\documentclass[a4paper]{llncs}
\usepackage{hareng}

\pdfoutput=1

\newif\iflong\longfalse

\hypersetup{%
pdftitle={Kleene Algebra with Tests and Coq Tools for While Programs},%
pdfauthor={Damien Pous},%
pdfkeywords={Coq proof assistant, reflexive tactic, Kleene algebra, KAT, finite automata, decision procedure, regular expressions},%
}

\title{Kleene Algebra with Tests \\
  and Coq Tools for While Programs} %
\author{Damien Pous} %
\institute{CNRS -- LIP, ENS Lyon, UMR 5668%
} %

\begin{document}

\maketitle

\begin{abstract}
  We present a Coq library about Kleene algebra with tests, including
  a proof of their completeness over the appropriate notion of
  languages, a decision procedure for their equational theory, and
  tools for exploiting hypotheses of a certain kind in such a
  theory.

  Kleene algebra with tests make it possible to represent if-then-else
  statements and while loops in most imperative programming
  languages. They were actually introduced by Kozen as an alternative
  to propositional Hoare logic.

  We show how to exploit the corresponding Coq tools in the context of
  program verification by proving equivalences of while programs,
  correctness of some standard compiler optimisations, Hoare rules for
  partial correctness, and a particularly challenging equivalence of
  flowchart schemes.
\end{abstract}

\section*{Introduction}
\label{sec:intro}

Kleene algebra with tests (KAT) have been introduced by
Kozen~\cite{kozen:97:kat}, as an equational system for program
verification. A Kleene algebra with tests is a Kleene algebra (KA)
with an embedded Boolean algebra of tests.  The Kleene algebra
component deals with the control-flow graph of the
programs---sequential composition, iteration, and branching---while
the Boolean algebra component deals with the conditions appearing in
if-then-else statements, while loops, or pre- and post-assertions.

This formalism is both concise and expressive, which allowed Kozen
and others to give detailed paper proofs about various problems in
program verification (see,
e.g.,~\cite{kozen:97:kat,kozenp00:kat:compiler:opts,kozen00:kat:hoare,angusk01:kat:schemato}).
More importantly, the equational theory of KAT is decidable and
complete over relational
models~\cite{kozens96:kat:completeness:decidability}, and hypotheses
of a certain kind can moreover be
eliminated~\cite{cohen94:ka:hypotheses,hardink02:kat:hypotheses}.
This suggests that a proof using KAT should not be done manually, but
with the help of a computer. The goal of the present work is to give
this possibility, inside the Coq proof assistant.

\medskip

The underlying decision procedure cannot be formulated, a priori, as a
simple rewriting system: it involves automata algorithms, it cannot be
defined in Ltac, at the meta-level, and it does not produce a
certificate which could easily be checked in Coq, a posteriori. This
leaves us with only one possibility: defining a reflexive
tactic~\cite{BoyerMoore81,AllenCHA90,GregoireM05}. 
Doing so is quite challenging: we basically have to prove completeness
of KAT axioms w.r.t.\ the model of guarded string languages (the
natural generalisation of languages for KA, to KAT), and to provide a
provably correct algorithm for language equivalence of KAT
expressions.

The completeness theorem is far from trivial; we actually have to
formalise a lot of preliminary material: finite sums, finite sets,
unique decomposition of Boolean expressions into sums of atoms,
regular expression derivatives, expansion theorem for regular
expressions, matrices, automata\dots{} As a consequence, we only give
here a high-level overview of the involved mathematics, leaving aside
standard definitions, technical details, or secondary formalisation
tricks. The interested reader can consult the library, which is
documented~\cite{pous:coq:ra}.

\paragraph{Outline.} We first present KAT and its models
(§\ref{sec:kat}). We then sketch the completeness proof
(§\ref{sec:completeness}), the decision procedure
(§\ref{sec:decision}), and the method used to eliminate hypotheses
(§\ref{sec:helim}). We finally illustrate the benefits of our tactics
on several case-studies (§\ref{sec:apps}), before discussing related
works (§\ref{sec:related:works}), and concluding (§\ref{ccl}).

\section{Kleene Algebra with Tests}
\label{sec:kat}

A Kleene algebra with tests consists of:
\begin{itemize}
\item a Kleene algebra
  $\tuple{X,{\cdot},{+},{\cdot^\star},1,0}$~\cite{kozen94:ka:completeness},
  i.e., an idempotent semiring with a unary operation, called ``Kleene
  star'', satisfying an axiom: $1+x\cdot{}x^\star \le x^\star$ and two
  inference rules: $y\cdot{}x \le x$ entails $y^\star\cdot{}x \le x$
  and the symmetric one.  (The preorder $(\le)$ being defined by $x\le y
  ~ \eqdef ~ x+y = y$.)
\item a Boolean algebra $\tuple{B,{\land},{\lor},{\neg},\top,\bot}$;
\item a homomorphism from $\tuple{B,{\land},{\lor},\top,\bot}$ to
  $\tuple{X,{\cdot},{+},1,0}$, that is, a function $[\cdot]:B\to X$
  such that $[a\land b]=[a]\cdot[b]$, $[a\lor b]=[a]+[b]$, $[\top]=1$,
  and $[\bot]=0$.
\end{itemize}
The elements of the set $B$ are called ``tests''; we denote them by
$a,b$. The elements of $X$, called ``Kleene elements'', are denoted by
$x,y,z$. We usually omit the operator ``$\cdot$'' from expressions,
writing $xy$ for $x\cdot y$. The following (in)equations illustrate the
kind of laws that hold in all Kleene algebra with tests:
\begin{mathpar}{}
  [ a\lor \neg a] = 1 \and{} %
  [ a\land(\neg a\lor b)] = [a][b] =
  [\neg(\neg a \lor \neg b)] \\
  x^\star x^\star = x^\star \and %
  (x+y)^\star = x^\star(yx^\star)^\star \and
  (x+x x y)^\star \le (x+x y)^\star\\  
  [a]([\neg a] x)^\star = [a] \and %
  [a]([a]x[\neg a]+[\neg a]y[a])^\star[a]
  \le (x y)^\star %
\end{mathpar}
The laws from the first line come from the Boolean algebra structure,
while the ones from the second line come from the Kleene algebra
structure. The two laws from the last line are more interesting: their
proof must mix both Boolean algebra and Kleene algebra
reasoning. They are left to the reader as a non-trivial exercice; the
tools we present in this paper allow one to prove them automatically.

\subsection{The model of binary relations}
\label{ssec:rel:model}

Binary relations form a Kleene algebra with tests; this is the main
model we are interested in, in practice. The Kleene elements are the
binary relations over a given set $S$, the tests are the predicates
over this set, and the star of a relation is its reflexive transitive
closure:
\begin{align*}
  \begin{aligned}
    X &= \pset{S\times S}\\
    x\cdot y &= \set{(p,q)\mid \exists r, (p,r)\in x \land (r,q)\in y}\\
    x+y &= x\cup y\\
    x^\star &= 
    \set{(p_0,p_n)\mid \exists p_1\dots p_{n-1}, \forall i<n, (p_i,p_{i+1})\in x}\\
    1 &= \set{(p,p)\mid p\in S}\\
    0 &= \emptyset %
    \hspace{3cm} [a] = \set{(p,p)\mid p\in a}
  \end{aligned}
  &&
  \begin{aligned}
    B &= \pset S\\
    a\land b &= a\cap b\\
    a\lor b &= a\cup b\\
    \neg a &= S\setminus a\\
    \top &= S\\
    \bot &= \emptyset
  \end{aligned}
\end{align*}
The laws of a Kleene algebra are easily proved for these operations;
note however that one needs either to restrict to decidable predicates
(i.e., to take \coqe$S -> bool$ or %
\coqe${p: S -> Prop | forall p, S p \/!S p}$ for $B$), or to assume
the law of excluded middle: $B$ must be a Boolean algebra, so that
negation has to be an involution. This choice for $B$ is left to the
user of the library.

This relational model is typically used to interpret imperative
programs: such programs are state transformers, i.e., binary relations
between states, and the conditions appearing in these programs are
just predicates on states. These conditions are usually decidable, so
that the above constraint is actually natural.

The equational theory of Kleene algebra with tests is complete over
the relational model~\cite{kozens96:kat:completeness:decidability}:
any equation $x=y$ that holds universally in this model can be proved
from the axioms of KAT. 
We do not need to formalise this theorem, but it is quite informative in
practice: by contrapositive, if an equation cannot be proved from KAT,
then it cannot be universally true on binary relations, meaning that
proving its validity for a particular instantiation of the variables
necessarily requires one to exploit additional properties of this
particular instance.

\subsection{Other models}
\label{ssec:other:models}

We describe two other models in the sequel: the syntactic model
(§\ref{ssec:syntactic:model}) and the model of guarded string
languages (§\ref{ssec:gsl:model}); these models have to be formalised
to build the reflexive tactic we aim at. 

There are other important models of KAT. First of all, any Kleene
algebra can be extended into a Kleene algebra with tests by embedding
the two-element Boolean lattice. We also have traces models (where one
keeps track of the whole execution traces of the programs rather than
just their starting and ending points), matrices over a Kleene algebra
with tests, but also models inherited from semirings like min-plus and
max-plus algebra%
. The latter
models have a degenerate Kleene star operation; they become useful
when one constructs matrices over them, for instance to study shortest
path algorithms.

Also note that like for Kleene
algebra~\cite{kozen98:ka:typed,pous:csl10:utas,bp:itp10:kacoq}, KAT
admits a natural ``typed'' generalisation, allowing for instance to
encompass heterogeneous binary relations and rectangular matrices. Our
Coq library is actually based on this generalisation, and this deeply
impacts the whole infrastructure; we however omit the corresponding
details and technicalities here, for the sake of clarity.

\subsection{KAT expressions}
\label{ssec:syntactic:model}

Let $p,q$ range over a set $\Sigma$ of \emph{letters} (or
\emph{actions}), and let $a_1,\dots,a_n$ be the elements of a finite
set $\Theta$ of \emph{primitive tests}. \emph{Boolean expressions} and
\emph{KAT expressions} are defined by the following syntax:
\begin{align*}
  \tag{Boolean expressions}
  a,b &::= a_i\in \Theta \mid a\land a \mid a\lor a \mid \neg a \mid
  \top \mid \bot\\
  \tag{KAT expressions}
  x,y &::= p\in \Sigma \mid [a] \mid x\cdot y \mid x+y \mid
  x^\star \mid 1 \mid 0\enspace.
\end{align*}
Given a Kleene algebra with tests $\mathcal K=\tuple{X,B,[\cdot]}$,
any pair of maps $\theta: \Theta\to B$ and $\sigma: \Sigma\to X$ gives
rise to a KAT homomorphism allowing to interpret expressions in
$\mathcal K$.
Given two such expressions $x$ and $y$, the equation $x=y$ is a
\emph{KAT theorem}, written $\KAT\vdash x=y$, when the equation holds
in any Kleene algebra with tests, under any interpretation. One checks
easily that KAT expressions quotiented by the latter relation form a
Kleene algebra with tests; this is the free Kleene algebra with tests
over $\Sigma$ and $\Theta$. (We actually use this impredicative
encoding of KAT derivability in the Coq library.)


\subsection{Guarded strings languages}
\label{ssec:gsl:model}

Guarded string languages are the natural generalisation of string
languages for Kleene algebra with tests. We briefly define them.

An \emph{atom} is a function from elementary tests $(\Theta)$ to
Booleans; it indicates which of these tests are satisfied. We let
$\alpha,\beta$ range over atoms, the set of which is denoted by $At$.
(Technically, we represent elementary tests as finite ordinals of a
given size $n$ $(\Theta=\mathtt{ord}~n)$, and we encode atoms as
ordinals $(\mathtt{At=ord}~2^n)$. This allows us to avoid functional
extensionality problems.) We let $u,v$ range over \emph{guarded
  strings}: 
alternating sequences of atoms and letters, which both start and end
with an atom:
\begin{align*}
  \alpha_1,p_1,\dots,\alpha_n,p_n,\alpha_{n+1}\enspace.
\end{align*}

The concatenation $u\ast v$ of two guarded strings $u,v$ is a partial
operation: it is defined only if the last atom of $u$ is equal to the
first atom of $v$; it consists in concatenating the two sequences and
removing the shared atom in the middle.
%

The Kleene algebra with tests of guarded string languages is obtained
by considering sets of guarded strings for $X$ and sets of atoms for $B$:
\begin{align*}
  \begin{aligned}
    X &= \pset{(At\times\Sigma)^\star\times At}\\
    x\cdot y &= \set{u\ast v \mid u\in x \land v\in y}\\
    x+y &= x\cup y\\
    x^\star &= 
    \set{u_1\ast \dots \ast u_n \mid \exists u_1\dots u_n, \forall i\leq n, u_i\in x}\\
    1 &= \set{\alpha\mid \alpha\in At}\\
    0 &= \emptyset %
    \hspace{3cm} [a] = \set{\alpha\mid \alpha\in a}
  \end{aligned}
  &&
  \begin{aligned}
    B &= \pset{At}\\
    a\land b &= a\cap b\\
    a\lor b &= a\cup b\\
    \neg a &= At\setminus a\\
    \top &= At\\
    \bot &= \emptyset
  \end{aligned}
\end{align*}
Note that we slightly abuse notation by letting $\alpha$ denote
either an atom, or a guarded string reduced to an atom.
Also note that the set $B=\pset{At}$ has to be represented by the Coq
type $\mathtt{At\to bool}$, to get a Boolean algebra on it.


\section{Completeness}
\label{sec:completeness}

Let $G$ be the unique homomorphism from KAT expressions to guarded
string languages such that
\begin{align*}
  G(a_i) &= \set{\alpha \mid \alpha(a_i)\text{ is true}} &
  G(p) &= \set{\alpha p\beta \mid \alpha,\beta\in At}
\end{align*}
Completeness of KAT over guarded string languages can be stated as
follows.
\begin{theorem}
  \label{thm:kat:compl}
  For all KAT expressions $x,y$, $G(x)=G(y)$ entails $\KAT\vdash x=y$.
\end{theorem}
This theorem is central to our development: it allows us to prove
(in)equations in arbitrary models of KAT, by resorting to an algorithm
deciding guarded string language equivalence (to be described in
§\ref{sec:decision}).

We closely follow Kozen and Smith'
proof~\cite{kozens96:kat:completeness:decidability}. This proof relies
on the completeness of Kleene algebra over languages, which we thus
need to prove first.

\subsection{Completeness of Kleene algebra axioms}
\label{ssec:ka:completeness}

Let $R$ be the Kleene algebra homomorphism from regular expressions to
(plain) string languages mapping a letter $p$ to the language
consisting of the single-letter word $p$. KA completeness over
languages can be stated as follows~\cite{kozen94:ka:completeness}:
\begin{theorem}
  \label{thm:ka:compl}
  For all regular expressions $x,y$, $R(x)=R(y)$ entails $\KA\vdash
  x=y$.
\end{theorem}
(Like for KAT, the judgement $\KA\vdash x=y$ means that $x=y$ holds in
any Kleene algebra, under any interpretation.) We already presented a
Coq formalisation of this theorem~\cite{bp:itp10:kacoq}, but our
development was over-complicated. We re-proved it from scratch here,
following a simpler path which we now describe.

The main idea of Kozen's proof consists in replaying automata
algorithms algebraically, using matrices to encode automata. The key
insight that allowed us to considerably simplify the corresponding
formalisation is that the algorithm used for this proof need not be
the same as the one to be executed by the reflexive tactic we
eventually define.
Indeed, we can take the simplest possible algorithm to prove KA
completeness, ignoring all complexity aspects, thus allowing us to
focus on conciseness and mathematical simplicity. 
In contrast, the algorithm to be executed by the final reflexive
tactic should be relatively efficient, but we do not need to prove it
complete, nor to replay its correctness algebraically: we only need
to prove its correctness w.r.t.\ languages, which is much
easier. 

\medskip

A preliminary step for the proof consists in proving that matrices
over a Kleene algebra form a Kleene algebra. The Kleene star for
matrices is non-trivial to define and to prove correct, but this can
be done with a reasonable amount of efforts once appropriate lemmas
and tools for block matrices have been set up.

A finite automaton can then be represented using three matrices
$(u,M,v)$ over regular expressions, where $u$ is a $(1,n)$-matrix, $M$
is a $(n,n)$-matrix, and $v$ is a $(n,1)$-matrix, $n$ being the number
of states of the automaton. Such a ``matricial automaton'' can be
evaluated into a regular expression by taking the product $u\cdot
M^\star\cdot v$, which is a scalar. The various classes of automata
can be recovered by imposing conditions on the coefficients of the
three matrices. For instance, a non-deterministic finite automaton
(NFA) is such that $u$ and $v$ are 01-vectors and the coefficients of
$M$ are sums of letters.

\medskip

Given a regular expression $x$, we construct a deterministic finite
automaton (DFA) $(u,M,v)$ such that $\KA\vdash x = uM^\star v$, as
follows.
\begin{enumerate}
\item First construct a NFA with epsilon transitions $(u'',M'',v'')$,
  such that $\KA\vdash x=u''M''^\star v''$. This is easily done by
  induction on $x$, using Thompson construction~\cite{thompson68}
  (which is compositional, unlike the construction we used
  in~\cite{bp:itp10:kacoq}).
\item Remove epsilon transitions to obtain a NFA $(u',M',v')$ such
  that $\KA\vdash u''M''^\star v'' = u'M'^\star v'$. We do it purely
  algebraically, in one line. In particular the transitive closure of
  epsilon transitions is computed using Kleene star on
  matrices. (Unlike in~\cite{bp:itp10:kacoq} we do not need a
  dedicated algorithm for this.)
\item Use the powerset construction to convert this NFA into a DFA
  $(u,M,v)$ such that $\KA\vdash u'M'^\star v' = uM^\star
  v$. Again, this is done algebraically, and we do not need to
  perform the standard `accessible subsets' optimisation.
\end{enumerate}
We can prove that for any DFA $(u,M,v)$, $R(uM^\star v)$ is the
language recognised by the DFA. Therefore, to obtain
Theorem~\ref{thm:ka:compl}, it suffices to prove that if two DFA
$(u,M,v)$ and $(s,N,t)$ recognise the same language, then $\KA\vdash
uM^\star v = sN^\star t$.
For this last step, it suffices to exhibit a Boolean matrix that
relates exactly those states of the two DFA that recognise the same
language. We need for that an algorithm to check language equivalence
of DFA states
; we reduce the problem to DFA emptiness, and we perform a simple
reachability analysis.

\medskip

All in all, the KA completeness proof itself only requires us 124 lines
of specifications, and 119 lines of proofs (according to \texttt{coqwc}).

\subsection{Completeness of KAT axioms}
\label{ssec:kat:completeness}

To obtain KAT completeness (Theorem~\ref{thm:kat:compl}), Kozen and
Smith~\cite{kozens96:kat:completeness:decidability} define a function
$\hat\cdot$ on KAT expressions that expands the expressions in such a
way that we have $\KAT\vdash x=y$ iff $\KA\vdash\hat x = \hat y$.
While this function can be thought as a reduction of KAT to KA, it
cannot be used in practice: it produces expressions that are almost
systematically exponentially larger than the given ones. It is however
sufficient to establish completeness; as explained earlier, we defer
actual computations to a completely different algorithm
(§\ref{sec:decision}).

More precisely, the function $\hat\cdot$ is defined in such a way
that we have:
\begin{align}
  \label{eq:hat:i}\KAT\vdash \hat x = x \tag{i} \\
  \label{eq:hat:ii} G(\hat x) = R(\hat x) \tag{ii}
\end{align}
We deduce KAT completeness as follows:
\begin{align*}
  & G(x) = G(y)\\
  \Leftrightarrow~ & G(\hat x) = G(\hat y) \tag{$G$ is a KAT morphism, and~\eqref{eq:hat:i}}\\
  \Leftrightarrow~ & R(\hat x) = R(\hat y) \tag{by~\eqref{eq:hat:ii}}\\
  \Rightarrow~ & \KA\vdash \hat x = \hat y \tag{KA completeness}\\
  \Rightarrow~ & \KAT\vdash \hat x = \hat y \tag{any KAT is a KA}\\
  \Leftrightarrow~ & \KAT\vdash x = y \tag{by~\eqref{eq:hat:i}}
\end{align*}
(Note that the last equation entails the first one, so that all these
statements are in fact equivalent.)

\medskip

The function $\hat\cdot$ is defined recursively over KAT expressions,
using an intermediate datastructure: formal sums of \emph{externally
  guarded terms} (i.e., either an atom, or a product of the form
$\alpha x \beta$). The case of a starred expression $x^\star$ is quite
involved: $\hat{x^\star}$ is defined by an internal recursion on the
length of the formal sum corresponding to $\hat x$.
The proof of the first equation~\eqref{eq:hat:i} is not too difficult
to formalise, using appropriate tools for finite sums (i.e., a
simplified form of big operators~\cite{BertotGBP08}, which we actually
use a lot in the whole development). The second one~\eqref{eq:hat:ii}
is more cumbersome, notably because we must deal with the two implicit
coercions appearing in its statement: formally, it has to be stated as
follows:
\begin{align*}
  i(G(\hat x)) = R(j(\hat x))\enspace, 
\end{align*}
where $i$ takes a guarded string language and returns a finite word
language on the alphabet $\Sigma\uplus\Theta\uplus\Theta$, and $j$
takes a KAT expression and returns a regular expression over this
extended alphabet, by pushing all negations to the leaves.

Apart from the properties of these coercion functions, the proof
of~\eqref{eq:hat:ii} mainly consists in rather technical arguments
about regular and guarded string languages concatenation.
All in all, once KA completeness has been proved, KAT completeness
requires us 278 lines of specifications, and 360 lines of proofs.

\section{Decision procedure}
\label{sec:decision}

To check whether two expressions denote the same language of guarded
strings, we use an algorithm based on a notion of \emph{partial
  derivatives} for KAT expressions.
Derivatives were introduced by Brzozowski~\cite{Brzozowski64} for
regular expressions; they make it possible to define a deterministic
automaton where the states of the automaton are the regular
expressions themselves. 

Derivatives can be extended to KAT expressions in a very natural
way~\cite{kozen08:kat:coalgebra}: we first define a Boolean function
$\epsilon_\alpha$, that indicates whether an expression accepts the
single atom $\alpha$; this function is then used to define the
derivation function $\delta_{\alpha,p}$, that intuitively returns what
remains of the given expression after reading the atom $\alpha$ and
the letter $p$. 
These two functions make it possible to give a coalgebraic
characterisation of the function $G$, which underpins the
correctness of the algorithm we sketch below:
\begin{align*}
  G(x)(\alpha) &= \epsilon_\alpha(x) &
  G(x)(\alpha\,p\,u) &= G(\delta_{\alpha,p}(x))(u)\enspace.
\end{align*}

Like with standard regular expressions, the set of derivatives of a
given KAT expression (i.e., the set of expressions that can be
obtained by repeatedly deriving w.r.t. arbitrary atoms and letters)
can be infinite. To recover finiteness, we switch to \emph{partial}
derivatives~\cite{Antimirov96}. Their generalisation to KAT should be
folklore; we define them in Fig.~\ref{fig:kat:pderiv}.
\begin{figure}[t]
  \centering
  \begin{align*}
    \begin{aligned}
      \delta'_{\alpha,p}(x+y) &= \delta'_{\alpha,p}(x)\cup \delta'_{\alpha,p}(y) \\
      \delta'_{\alpha,p}(x y) &=
      \begin{cases}
        \delta'_{\alpha,p}(x) y \cup
        \delta'_{\alpha,p}(y)&\text{if }\epsilon_\alpha(x)\\
        \delta'_{\alpha,p}(x) y &\text{otherwise}
      \end{cases}\\
      \delta'_{\alpha,p}(x^\star) &= \delta'_{\alpha,p}(x) x^\star\\
    \end{aligned}&&\qquad
    \begin{aligned}
      \delta'_{\alpha,p}(q) &=
      \begin{cases}
        \set{1} & \text{if }p=q\\
        \emptyset & \text{otherwise}
      \end{cases}\\
      \delta'_{\alpha,p}([a]) &= \emptyset\\\\
    \end{aligned}
  \end{align*}
  \vspace{-1em}
  \caption{Partial derivatives for KAT expressions}
\label{fig:kat:pderiv}
\end{figure}
We use the notation $Xy$ to denote the set $\set{xy\mid x\in X}$ when
$X$ is a set of expressions and $y$ is an expression. The partial
derivation function $\delta'_{\alpha,p}$ returns a (finite) set of
expressions rather than a single one; this corresponds to the fact
that we build a non-deterministic automaton. Still abusing notations,
by letting a set of expressions denote the sum of its elements,
we prove that $\KAT\vdash\delta_{\alpha,p}(x) =
\delta'_{\alpha,p}(x)$.

\medskip

Now call \emph{bisimulation} any relation $R$ between sets of
expressions such that whenever $X\mathrel R Y$, we have
\begin{itemize}
\item $\epsilon(X) = \epsilon(Y)$ and
\item $\forall\alpha\in At, \forall p\in\Sigma$,
  $\delta'_{\alpha,p}(X)\mathrel R\delta'_{\alpha,p}(Y)$.
\end{itemize}
We show that if there is a bisimulation $R$ such that $X\mathrel R
Y$, then $G(X)=G(Y)$ (the converse also holds).  This gives us an
algorithm to decide language equivalence of two KAT expressions $x,y$:
it suffices to try to construct a bisimulation that relates the
singletons $\set x$ and $\set y$. 
This algorithm terminates because the set of partial derivatives
reachable from a pair of expressions is finite (we do not need to
formalise this fact since we just need the correctness of this
algorithm).

There is a lot of room for optimisation in our implementation---for
instance, we use unordered lists to represent binary relations. An
important point in our design is that such optimisations can be
introduced and proved correct independently from the completeness
proof for KAT, which gives us much more flexibility than in our
previous work on Kleene algebra~\cite{bp:itp10:kacoq}.

\subsection{Building a reflexive tactic}
\label{ssec:kat:tac}

Using standard
methodology~\cite{BoyerMoore81,AllenCHA90,GregoireM05}, 
we finally pack the previous ingredients into a Coq reflexive tactic
called \code{kat}, allowing us to close automatically any goal which
belongs to the equational theory of KAT.

The tactic works on any model of KAT: those already declared in the
library (relations, languages, matrices, traces), but also the ones
declared by the user. 
%
The reification code is written in OCaml; it is quite complicated for
at least two reasons: KAT is a two-sorted structure, and we actually
deal with ``typed'' KAT, as explained in §\ref{ssec:other:models},
which requires us to work with a dependently typed syntax.

For the sake of simplicity, the Coq algorithm we implemented for KAT
does not produce a counter-example in case of failure. To be able to
give such a counter-example to the user, we actually run an OCaml
copy of the algorithm first (extracted from Coq, and modified by hand
to produce counter-examples). This has two advantages: the tactic is
faster in case of failure, and the counter-example---a guarded
string---can be pretty-printed in a nicer way.

\section{Eliminating hypotheses}
\label{sec:helim}

The above \code{kat} tactic works for the equational theory of KAT,
i.e., the (in)equations that hold in any model of KAT, under any
interpretation.  In particular, this tactic does not make use of any
hypothesis which is specific to the model or to the interpretation.
Some hypotheses can however be
exploited~\cite{cohen94:ka:hypotheses,hardink02:kat:hypotheses}: those
having one of the following shapes.
\begin{enumerate}[(i)]
\item\label{sh:hoare} $x = 0$;
\item\label{sh:appb} $[a]x = x[b]$, $[a]x \le x[b]$, or $x[b] \le [a]x$;
\item\label{sh:ppb} $x \le [a]x$ or $x \le x[a]$
\item\label{sh:ab} $a = b$ or $a \le b$;
\item\label{sh:cpc} $[a]p = [a]$ or $p[a] = [a]$, for atomic $p$ $(p\in\Sigma)$;
\end{enumerate}

Equations of the first kind~\eqref{sh:hoare} are called ``Hoare''
equations, for reasons to become apparent in
§\ref{ssec:hoare:logic}. They can be eliminated using the following
implication:
\begin{align}
  \label{eq:elim:hoare}
  \tag{$\dag$}
  \begin{cases}
    x+uzu = y+uzu \\ z=0
  \end{cases}
  &\quad\text{entails}\quad x=y\enspace.
\end{align}
This implication is valid for any term $u$, and the method is
complete~\cite{hardink02:kat:hypotheses} when $u$ is taken to be the
universal KAT expression, $\Sigma^\star$.
Intuitively, for this choice of $u$, $uzu$ recognizes all guarded
strings that contain a guarded string of $z$ as a
substring. Therefore, when checking that $x+uzu = y+uzu$ are language
equivalent rather than $x=y$, we rule out all counter-examples to
$x=y$ that contain a substring belonging to $z$: such counter-examples
are irrelevant since $z$ is known to be empty.

\medskip

Equations of the shape~\eqref{sh:ppb} and~\eqref{sh:ab} are actually
special cases of those of the shape~\eqref{sh:appb}, which are in turn
equivalent to Hoare equations. For instance, we have $[a]x \le x[b]$
iff $[a]x[\neg b] = 0$.
%
Moreover, two hypotheses of shape~\eqref{sh:hoare} can be merged into
a single one using the fact that $x=0 \land y=0$ iff
$x+y=0$. Therefore, we can aggregate all hypotheses of
shape~(\ref{sh:hoare}-\ref{sh:ab}) into a single one (of
shape~\eqref{sh:hoare}), and use the above technique just once.

\medskip

\noindent
Hypotheses of shape~\eqref{sh:cpc} are handled differently, using the following
equivalence:
\begin{align}
  \label{eq:elim:cpc}
  \tag{$\ddag$}
  [a]p = [a] &\quad\text{iff}\quad p = [\neg a]p+[a]\enspace,
\end{align}
This equivalence allows us to substitute $[\neg a]p+[a]$ for $p$ in
the considered goal---whence the need for $p$ to be atomic. Again, the
method is complete~\cite{hardink02:kat:hypotheses}, i.e.,
\begin{align*}
  \KAT\vdash ([a]p = [a] \Rightarrow x=y) &\quad\text{iff}\quad
  \KAT\vdash x\theta=y\theta \tag{$\theta=\set{p\mapsto[\neg a]p+[a]}$}
\end{align*}

\subsection{Automating elimination of hypotheses in Coq}
\label{ssec:hkat}

The previous techniques to eliminate some hypotheses in KAT can be
easily automated in Coq. We first prove once and for all the
appropriate equivalences and implications (the tactic \code{kat} is
useful for that). We then define some tactics in Ltac that collect
hypotheses of shape~(\ref{sh:hoare}-\ref{sh:ab}), put them into
shape~\eqref{sh:hoare}, and aggregate them into a single one which is
finally used to update the goal according
to~\eqref{eq:elim:hoare}. Separately, we define a tactic that rewrites
in the goal using all hypotheses of shape~\eqref{sh:cpc},
through~\eqref{eq:elim:cpc}. Finally, we obtain a tactic called
\code{hkat}, that just preprocesses the conclusion of the goal using
all hypotheses of shape~(\ref{sh:hoare}-\ref{sh:cpc}) and then calls
the \code{kat} tactic.
Note that the completeness of this
method~\cite{hardink02:kat:hypotheses} is a meta-theorem; we do not
need to formalise it.

\section{Case studies}
\label{sec:apps}

We now present some examples of Coq formalisations where one can take
advantage of our library.

\subsection{Bigstep semantics of `while' programs}
\label{ssec:imp}

The bigstep semantics of `while' programs is teached in almost any
course on semantics and programming languages. Such programs can be
embedded into KAT in a straightforward way~\cite{kozen00:kat:hoare},
thus providing us with proper tools to reason about them. Let us
formalise such a language in Coq.

Assume a type \code{state} of states, a type \code{loc} of memory
locations, and an \code{update} function allowing to update the value
of a memory location. Call \emph{arithmetic expression} any function
from states to natural numbers, and \emph{Boolean expression} any
function from states to Booleans (we use a partially shallow
embedding). The `while' programming language is defined by the
inductive type below:

\medskip
\begin{twolistings}
\begin{coq}
Variable loc, state: Set.
Variable update: loc -> nat -> state -> state.
 
Definition expr := state -> nat.
Definition test := state -> bool. 
\end{coq}&
\begin{coq}
Inductive prog :=
| skp
| aff (l: loc) (e: expr)
| seq (p q: prog)
| ite (b: test) (p q: prog)
| whl (b: test) (p: prog).
\end{coq}
\end{twolistings}
\medskip

\noindent
The bigstep semantics of such programs is given as a ``state
transformer'', i.e., a binary relation between states. 
Following standard textbooks, one can define this semantics in Coq
using an inductive predicate:

\begin{coq}
Inductive bstep: prog -> rel state state :=
| s_skp: \forall s, bstep skp s s
| s_aff: \forall l e s, bstep (aff l e) s (update l (e s) s)
| s_seq: \forall p q s s' s'', bstep p s s' -> bstep q s' s'' -> bstep (seq p q) s s''
| s_ite_ff: \forall b p q s s', ! b s -> bstep q s s' -> bstep (ite b p q) s s'
| s_ite_tt: \forall b p q s s', b s -> bstep p s s' -> bstep (ite b p q) s s'
| s_whl_ff: \forall b p s, ! b s -> bstep (whl b p) s s
| s_whl_tt: \forall b p s s', b s -> bstep (seq p (whl b p)) s s' -> bstep (whl b p) s s'.
\end{coq}

\noindent
Alternatively, one can define this semantic through the relational
model of KAT, by induction over the program structure:

\begin{coq}
Fixpoint bstep (p: prog): rel state state :=
  match p with
    | skp => 1
    | seq p q => bstep p;bstep q
    | aff l e => upd l e
    | ite b p q => [b];bstep p+[!b];bstep q
    | whl b p => ([b];bstep p)^*;[!b]
  end.
\end{coq}
(Notations come for free since binary relations are already declared
as a model of KAT in our library.)
The `skip' instruction is interpreted as the identity relation;
sequential composition is interpreted by relational composition.
Assignments are interpreted using an auxiliary function, defined as
follows:
\begin{coq}
Definition upd l e: rel state state := fun s s' => s' = update l (e s) s.
\end{coq}
For the `if-then-else' statement, the Boolean expression \code{b} is a
predicate on states, i.e., a test in our relational model of KAT; this
test is used to guard both branches of the possible execution paths.
Accordingly for the `while' loop, we iterate the body of the loop
guarded by the test, using Kleene star. We make sure one cannot exit
the loop before the condition gets false by post-guarding the
iteration with the negation of this test.

This alternative definition is easily proved equivalent to the
previous one. Its relative conciseness makes it easier to read; more
importantly, this definition allows us to exploit all theorems and
tactics about KAT, for free.
For instance, suppose that one wants to prove some program
equivalences. First define program equivalence, through the bigstep
semantics:
\begin{coq}
Notation "p ~ q" := (bstep p == bstep q). 
\end{coq}
(The ``\code{==}'' symbol denotes equality in the considered KAT
model; in this case, relational equality.)
The following lemmas about unfolding loops and dead code elimination,
can be proved automatically.
\begin{coq}
Lemma two_loops b p:[spc] whl b (whl b p) ~ whl b p.
Proof. simpl. kat. Qed.         
(* ([b];(([b];bstep p)^*;[!b]))^*;[!b] == ([b];bstep p)^*;[!b] *)
\end{coq}

\begin{coq}
Lemma fold_loop b p:[spc] whl b (p ;; ite b p skp) ~ whl b p.
Proof. simpl. kat. Qed.
(* ([b];(bstep p;([b];bstep p+[!b];1)))^*;[!b] == ([b];bstep p)^*;[!b] *)
\end{coq}

\begin{coq}
Lemma dead_code a b p q r: [spc] whl (a\/b) p ;; ite b q r  ~ [spc] whl (a\/b) p ;;[spc]r.
Proof. simpl. kat. Qed.    
(* ([a\/b];bstep p)^*;[!(a\/b)];([b];bstep q+[!b];bstep r) 
                                       == ([a\/b];bstep p)^*;[!(a\/b)];bstep r *)
\end{coq}
(The semicolon in program expressions is a notation for sequential
composition; the comments below each proof show the intermediate goal
where the \code{bstep} fixpoint has been simplified, thus revealing
the underlying KAT equality.)

Of course, the \code{kat} tactic cannot prove arbitrary program
equivalences: the theory of KAT only deals with the control-flow graph
of the programs and with the Boolean expressions, not with the concrete
meaning of assignments or arithmetic expressions.
We can however mix automatic steps with manual ones. Consider for
instance the following example, where we prove that an assignment can
be delayed. Our tactics cannot solve it automatically since some
reasoning about assignments is required; however, by asserting
manually a simple fact (in this case, an equation of
shape~\eqref{sh:appb}), the goal becomes provable by the \code{hkat}
tactic.

\begin{coq}
Definition subst l e (b: test): test := fun s => b (update l (e s) s).
Lemma aff_ite l e b p q: (l<-e;; ite b p q) ~ (ite (subst l e b) (l<-e;; p) (l<-e;; q)).
Proof.
  simpl. (* upd l e;([b];bstep p+[!b];bstep q) == 
         [subst l e b];(upd l e;bstep p);[!subst l e b];(upd l e;bstep q) *)
  assert (upd l e;[b] == [subst l e b];upd l e) by (cbv;; firstorder;; subst;; eauto).
  hkat.
Qed.
\end{coq}

\subsection{Hoare logic for partial correctness}
\label{ssec:hoare:logic}

Hoare logic for partial correctness~\cite{Hoare69} is subsumed by
KAT~\cite{kozen00:kat:hoare}. 
%
The key ingredient in Hoare logic is the notion of a ``Hoare triple''
$\hoare A p B$, where $p$ is a program, and $A,B$ are two formulas
about the memory manipulated by the program, respectively called pre-
and post-conditions. A Hoare triple $\hoare A p B$ is \emph{valid} if
whenever the program $p$ starts in some state $s$ satisfying $A$ and
terminates in a state $s'$, then $s'$ satisfies $B$. Such a statement
can be translated into KAT as a simple equation:
\begin{align*}
  [A]p[\neg B] = 0
\end{align*}
Indeed, $[A]p[\neg B]=0$ precisely means that there is no execution
path along $p$ that starts in $A$ and ends in $\neg B$.
Such equations are Hoare equations (they have the
shape~\eqref{sh:hoare} from §\ref{sec:helim}), so that they can be
eliminated automatically. As a consequence, inference rules of Hoare
logic can be proved automatically using the \code{hkat} tactic.
For instance, for the `while' rule, we get the following script:
\begin{coq}
Lemma rule_whl A b p: {A/\b} p {A} -> {A} whl b p {A/\!b}.
Proof. simpl. hkat. Qed.
(* [A/\b];bstep p;[!A] == 0 -> [A];(([b];bstep p)^*;[!b]);[!(A/\!b)] == 0 *)
\end{coq}

\subsection{Compiler optimisations}
\label{sec:compiler:opts}

Kozen and Patron~\cite{kozenp00:kat:compiler:opts} use KAT to verify a
rather large range of standard compiler optimisations, by equational
reasoning. Citing their abstract, they cover ``\emph{dead code
  elimination, common subexpression elimination, copy propagation,
  loop hoisting, induction variable elimination, instruction
  scheduling, algebraic simplification, loop unrolling, elimination of
  redundant instructions, array bounds check elimination, and
  introduction of sentinels}''. They cannot use automation, so that
the size of their proofs ranges from a few lines to half a page of KAT
computations.

We formalised all those equational proofs using our library. Most of
them can actually be solved instantaneously, by a simple call to the
\code{hkat} tactic.
For the few remaining ones, we gave three to four lines proofs,
consisting of first rewriting using hypotheses that cannot be
eliminated, and then a call to \code{hkat}.

The reason why \code{hkat} performs so well is that most assumptions
allowing to optimise the code in these examples are of the
shape~(\ref{sh:hoare}-\ref{sh:cpc}). For instance, to state that an
instruction $p$ has no effect when $[a]$ is satisfied, we use an
assumption $[a]p=[a]$. Similarly, to state that the execution of a
program $x$ systematically enforces $[a]$, we use an assumption
$x=x[a]$.
The assumptions that cannot be eliminated are typically those of the
shape $pq=qp$: ``the instructions $p$ and $q$ commute''; such
assumptions have to be used manually.


\subsection{Flowchart schemes}
\label{sec:program:schematology}

The last example we discuss here is due to Paterson, it consists in
proving the equivalence of two flowchart schemes (i.e., goto
programs---see Manna's book~\cite{Manna74} for a complete description
of this model). The two schemes are given in
Appendix~\ref{app:paterson}; Manna proves their equivalence using
several successive graph transormations. His proof is really
high-level and informal; it is one page long, plus three additional
pages to draw intermediate flowcharts schemes. Angus and
Kozen~\cite{angusk01:kat:schemato} give a rather detailed equational
proof in KAT, which is about six pages long. Using the \code{hkat}
tactic together with some ad-hoc rewriting tools, we managed to
formalise Angus and Kozen's proof in three rather sparse screens.

Like in Angus and Kozen's proof, we progressively modify the KAT
expression corresponding to the first schema, to make it evolve
towards the expression corresponding to the second schema.
Our mechanised proof thus roughly consists in a sequence of
transitivity steps closed by \code{hkat}, allowing us to perform some
rewriting steps manually and to move to the next step. 
This is illustrated schematically by the code presented in
Fig.~\ref{fig:paterson}.
\begin{figure}[t]
  \centering
\begin{coq}
Lemma Paterson: x_1 == z.
Proof.
  transitivity y_1. hkat.         (* x_1 == y_1 *)
  a few rewriting steps transforming y_1 into x_2.
  transitivity y_2. hkat.         (* x_2 == y_2 *)
  a few rewriting steps transforming y_2 into x_3.
  (* ... *)
  transitivity y_19. hkat.        (* x_19 == y_19 *)
  a few rewriting steps transforming y_19 into x_20.
  hkat.                           (* x_20 == z *)
Qed.
\end{coq}\vspace{-1em}
  \caption{Squeleton for the proof of equivalence of Paterson's flowchart schems}
  \label{fig:paterson}
\end{figure}

Most of our transitivity steps (the $y_i$'s) already appear in Angus
and Kozen's proof; we can actually skip a lot of their steps, thanks
to \code{hkat}. Some of these simplifications can be spectacular: for
instance, they need one page to justify the passage between their
expressions (24) and (27), while a simple call to \code{hkat} does the
job; similarly for the page they need between their steps (38) and
(43).

\section{Related works}
\label{sec:related:works}

Several formalisations of algorithms and results related to regular
expressions and languages have been proposed since we released our Coq
reflexive decision procedure for Kleene algebra~\cite{bp:itp10:kacoq}:
partial derivatives for regular expressions~\cite{AlmeidaMPS10},
regular expression
equivalence~\cite{CoquandS11,KraussN12,Asperti12,MoreiraPS12}, regular
expression matching~\cite{Komendantsky12}. None of these works
contains a formalised proof of completeness for Kleene algebra, so
that they cannot be used to obtain a general tactic for KA (note
however that Krauss and Nipkow~\cite{KraussN12} obtain an Isabelle/HOL
tactic for binary relations using a nice trick to sidestep the
completeness proof---but they cannot deal with other models of KA).

On the more algebraic side, Struth et
al.~\cite{FosterS12,ArmstrongS12} showed how to formalise and use
relation algebra and Kleene algebra in Isabelle/HOL; they exploit the
automation tools provided by this assistant, but they do not try to
define decision procedures specific to Kleene algebra, and they do not
prove completeness.

To the best our knowledge, the only formalisation of KAT prior to the
present work is due to Pereira and Moreira~\cite{PereiraM08}, in Coq.
They state all axioms of KAT, derive some simple consequences of these
axioms (e.g., Boolean disjunction distribute over conjunction, Kleene
star is monotone), and use them to manually prove the inference rules
of Hoare logic, as we did automatically in
§\ref{ssec:hoare:logic}. They do not provide models, automation tools,
or completeness proof.

\section{Conclusion}
\label{ccl}

We presented a rather exhaustive Coq formalisation of Kleene algebra
with tests: axiomatisation, models, completeness proof, decision
procedure, elimination of hypotheses. We then showed several use-cases
for the corresponding library: proofs about while programs and Hoare
logic, certification of standard compiler optimisations, and
equivalence of flowchart schemes.

Most of the theoretical material is due to Kozen et
al.~\cite{kozen94:ka:completeness,kozens96:kat:completeness:decidability,kozen:97:kat,kozen98:ka:typed,kozenp00:kat:compiler:opts,kozen00:kat:hoare,angusk01:kat:schemato,hardink02:kat:hypotheses,kozen08:kat:coalgebra},
so that our contribution mostly lies in the Coq mechanisation of these
ideas.
The completeness proof was particularly challenging to formalise, and
lots of aspects of this work could not be explained in this extended
abstract: how to encode the algebraic hierarchy, how to work
efficiently with finite sets and finite sums, how to exploit symmetry
arguments, reflexive normalisation tactics, tactics about lattices,
finite ordinals and encodings of set-theoretic constructs in
ordinals\dots


The Coq library is available online~\cite{pous:coq:ra}; it is
documented and axiom-free; its overall structure is given in
Appendix~\ref{app:structure}. This library actually has a larger scope
than what we presented here: our long-term goal is to formalise and
automate other fragments of relation algebra (residuated structures,
Kleene algebra with converse, allegories\dots), so that the library is
designed to allow for such extensions. For instance normalisation
tactics and an ad-hoc semi-decision procedures are already defined for
algebraic structures beyond Kleene algebra and KAT.

According to \texttt{coqwc}, the library consists of 4377 lines of
specifications and 3020 lines of proofs, that distribute as follows.
Overall, this is slightly less than our previous library for
KA~\cite{bp:itp10:kacoq} (5105+4315 lines), and we do much more: not
only we handle KAT, but we also lay the ground for the mechanisation
of other fragments of relation algebra, as explained above.

\begin{center}
  \begin{tabular}{|l|r|r|r|}
    \hline
    &specifications&proofs&comments\\
    \hline
    ordinals, comparisons, finite sets\dots&674&323&225\\  
    algebraic hierarchy&490&374&216\\
    models (languages, relations, expressions\dots)&1279&461&404\\
    linear algebra, matrices&534&418&163\\
    completeness, decisions procedure, tactics&1400&1444&740\\
    \hline
  \end{tabular}
\end{center}

The resulting theorems and tactics allowed us to shorten significantly
a number of paper proofs---those about Hoare logic, compiler
optimisations, and flowchart schemes. Getting a way to guarantee that
such proofs are correct is important: although mathematically simple,
they tend to be hard to proofread (we invite the skeptical reader to
check Angus and Kozen's paper proof of Paterson
example~\cite{angusk01:kat:schemato}).  Moreover, automation greatly
helps when searching for such proofs: being able to get either a proof
or a counter-example for any proposed equation is a big plus: it makes
it much easier to progress in the overall proof.

\bibliographystyle{abbrv}
\bibliography{short,bib}

\clearpage
\appendix
\section{Paterson's flowchart schemes}
\label{app:paterson}

Here are the two flowchart schemes we proved equivalent
(§\ref{sec:program:schematology}), they appear
in~\cite[pages 254 and 258]{Manna74}.

\bigskip

\noindent
\includegraphics[width=.4\linewidth]{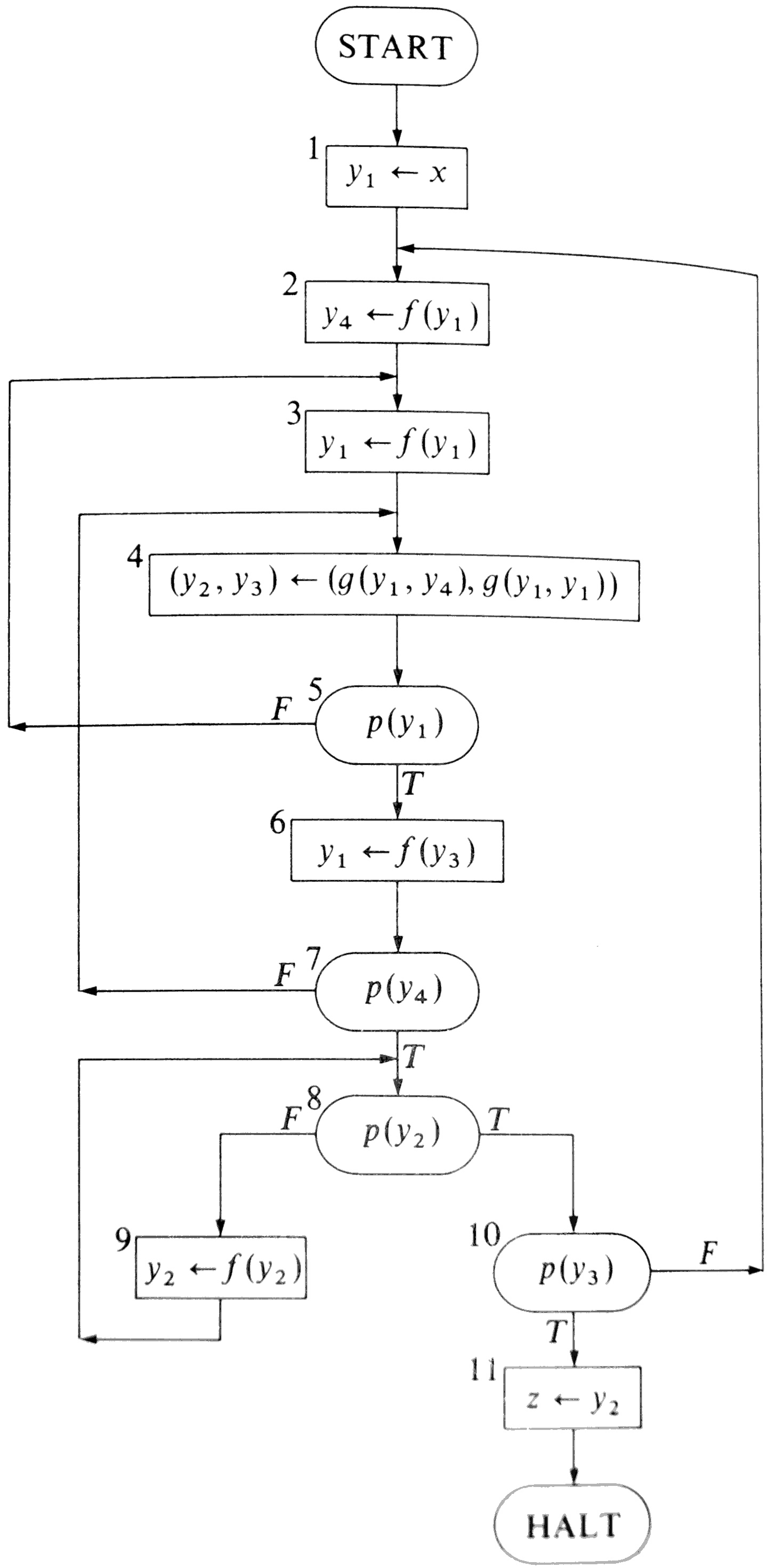}
\hfill
\includegraphics[width=.4\linewidth]{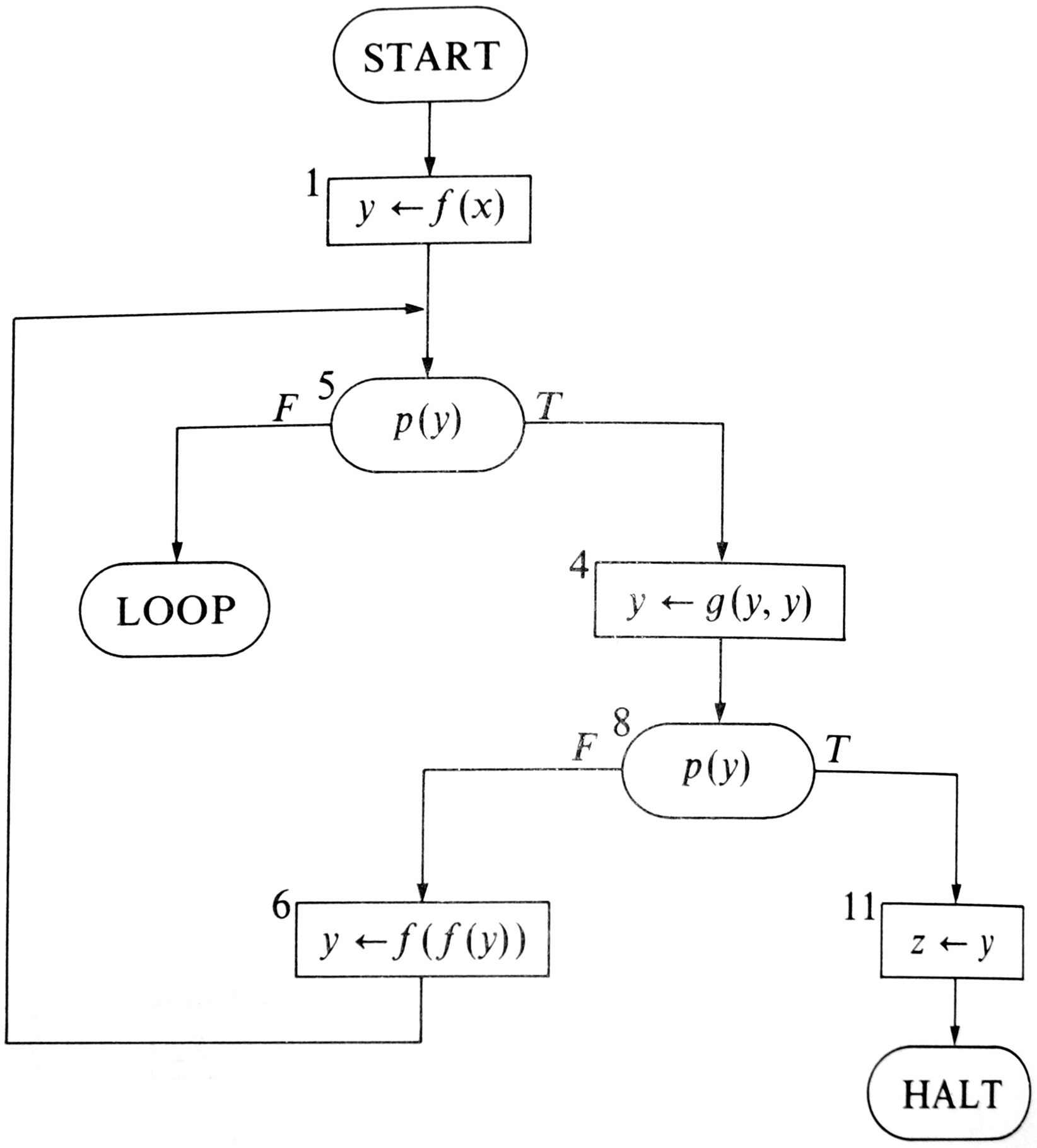}\\[.5em]
\mbox{}\hfill Schema $S_{6A}$ \hfill\hfill  Schema $S_{6E}$ \hfill\mbox{}\\

\smallskip\noindent
Following Angus and Kozen's notations~\cite{angusk01:kat:schemato},
these two schemes can be converted into the following KAT expressions:
\begin{align*}
  S_{6A} &= x_1 p_{41} p_{11} q_{214} q_{311}
  \left([\neg a_{1}] p_{11} q_{214} q_{311}\right)^\star [a_{1}] p_{13} \\
  &\quad \left(\left([\neg a_{4}]+[a_{4}]([\neg
      a_{2}]p_{22})^\star[a_{2}\land \neg a_{3}]p_{41}p_{11}\right)
    q_{214}q_{311} \left([\neg
      a_{1}]p_{11}q_{214}q_{311}\right)^\star[a_{1}]p_{13}\right)^\star\\
  &\quad [a_{4}]\left([\neg a_{2}]p_{22}\right)^\star[a_{2}\land a_{3}]z_{2} \\[.5em]
  S_{6E} &= s_1[a_1]q_1\left([\neg a_1]r_1[a_1]q_1\right)^\star[a_1]z_1\enspace,
\end{align*}
where the tests and actions are interpreted as follows:
\begin{align*}
  x_i &\eqdef y_i \leftarrow x &   
  z_i &\eqdef z \leftarrow y_i &
  a_i &\eqdef P(y_i) \\
  p_{ij} &\eqdef y_i \leftarrow f(y_j) &
  q_{ijk} &\eqdef y_i \leftarrow g(y_j,y_k)
\end{align*}
(Note that we actually renamed the local variable $y$ from schema
$S_{6E}$ into $y_1$, for the sake of uniformity.)

\clearpage
\section{Overall structure of the library}
\label{app:structure}

Here is a succinct description of each module from the library:

\newcommand\module[1]{\item \coqe$#1$:}
\begin{description}
  \item \setulcolor{blue}\ul{Utilities}
  \begin{description}
    \module{common} basic tactics and definitions used
    throughout the library
    \module{comparisons} types with decidable equality
    and ternary comparison function
    \module{positives} simple facts about binary positive numbers
    \module{ordinal} finite ordinals, finite sets of
    finite ordinals
    \module{pair} encoding pairs of ordinals as ordinals
    \module{powerfix} simple pseudo-fixpoint iterator
    \module{lset} sup-semilattice of finite sets represented as lists
  \end{description}
  \item \setulcolor{dkgreen}\ul{Algebraic hierarchy}
  \begin{description}
    \module{level} bitmasks allowing us to refer to an
    arbitrary point in the hierarchy
    \module{lattice} ``flat'' structures, from preorders
    to Boolean lattices
    \module{monoid} typed structures, from po-monoids to residuated Kleene lattices
    \module{kat} Kleene algebra with tests
    \module{kleene} Basic facts about Kleene algebra
    \module{normalisation} normalisation and semi-decision tactics for relation algebra
  \end{description}
  \item \setulcolor{yellow}\ul{Models}
    \begin{description}
    \module{prop} distributive lattice of propositions
    \module{boolean} Boolean trivial lattice, extended to a monoid.
    \module{rel} heterogeneous binary relations
    \module{lang} word languages
    \module{traces} trace languages
    \module{atoms} atoms of the free Boolean lattice over a finite set
    \module{glang} guarded string languages
    \module{lsyntax} free lattice (Boolean expressions)
    \module{syntax} free relation algebra
    \module{regex} regular expressions
    \module{gregex} KAT expressions (typed---for KAT completeness)
    \module{ugregex} untyped KAT expressions (untyped---for KAT decision procedure)
  \end{description}
  \item \setulcolor{orange}\ul{Untyping theorems}
  \begin{description}
    \module{untyping} untyping theorem for structures
    below KA with converse
    \module{kat_untyping} untyping theorem for guarded
    string languages
  \end{description}
  \item \setulcolor{red}\ul{Linear algebra}
  \begin{description}
    \module{sups} finite suprema/infima (a la bigop, from ssreflect)
    \module{sums} finite sums
    \module{matrix} matrices over all structures
    supporting this construction
    \module{matrix_ext} additional operations and
    properties about matrices
    \module{rmx} matrices of regular expressions
    \module{bmx} matrices of Booleans
  \end{description}
  \item \setulcolor{violet}\ul{Automata, completeness}
  \begin{description}
    \module{dfa} deterministic finite state automata, decidability of language inclusion
    \module{nfa} matricial non-deterministic finite state automata
    \module{ugregex_dec} decision of language equivalence for KAT expressions
    \module{ka_completeness} (untyped) completeness of Kleene algebra 
    \module{kat_completeness} (typed) completeness of Kleene
    algebra with tests
    \module{kat_reification} tools and definitions for KAT reification
    \module{kat_tac} decision tactics for KA and KAT,
    elimination of hypotheses
  \end{description}
\end{description}

\clearpage

Here are the dependencies between these modules:

\bigskip
\noindent\hspace{-.1\linewidth}
\includegraphics[width=1.2\linewidth]{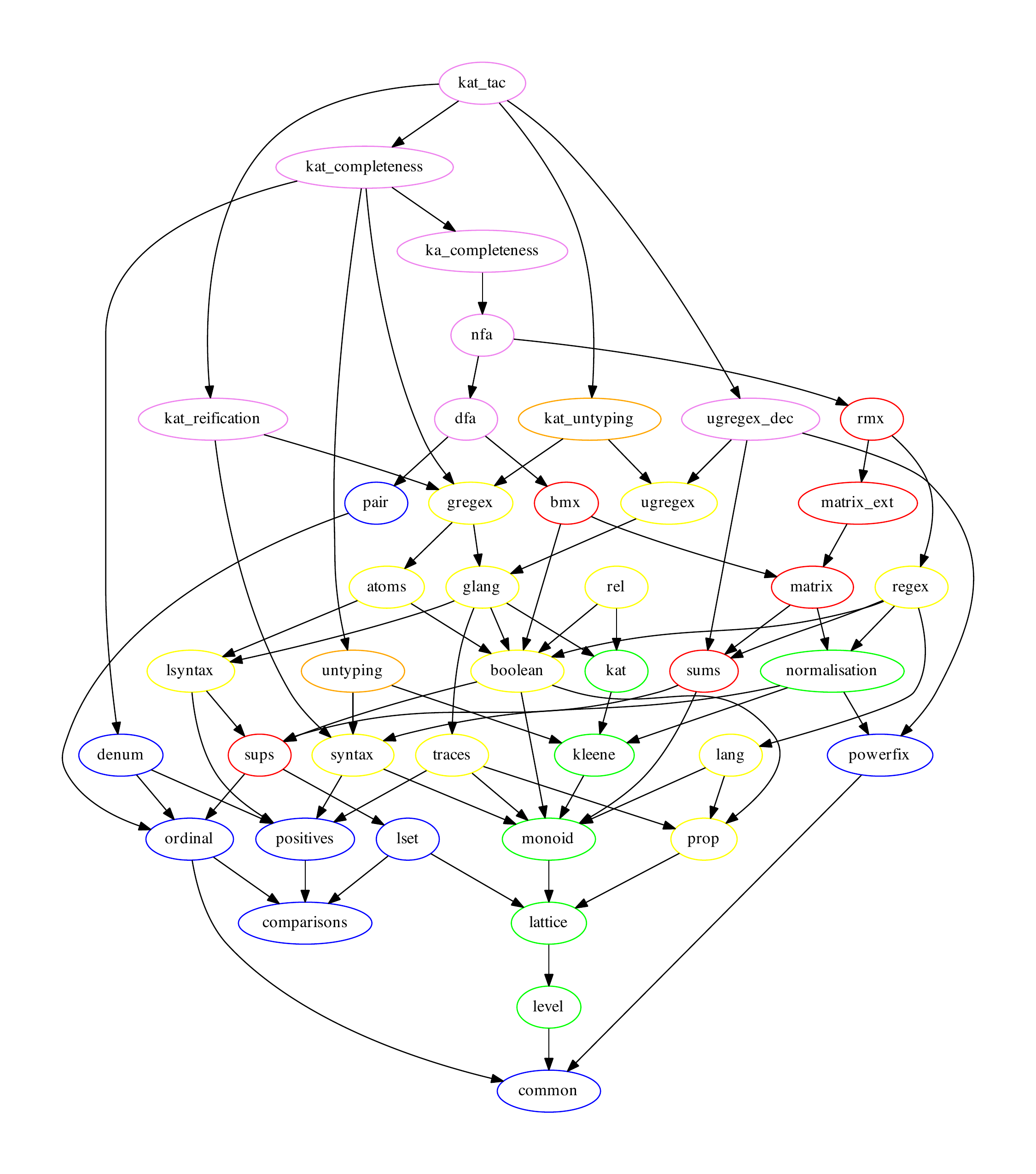}

\end{document}